\documentclass[usenatbib]{mn2e}
\usepackage{epsfig}
\usepackage{amsmath,amssymb}
\newcommand{\kpc}{{\,\rm kpc}}
\newcommand{\Mpc}{{\,\rm Mpc}}
\newcommand{\G}{{\rm G}}
\newcommand{\Om}{\Omega_m}
\newcommand{\Ol}{\Omega_\Lambda}
\newcommand{\sta}{_{\ast}}
\newcommand{\dis}{_{dis}}
\newcommand{\B}{_{\!B}}
\newcommand{\vir}{_{vir}}
\newcommand{\tot}{_{tot}}
\renewcommand{\vec}[1]{{\mathbf #1}}
\title{On the dynamics of the satellite galaxies in NGC 5044}
\author[A. Faltenbacher and W.G. Mathews]{
{\parbox[t]\textwidth{Andreas Faltenbacher
and 
William G. Mathews}} 
\vspace*{6pt} \\
UCO/Lick Observatory,
University of California at Santa Cruz,
1156 High Street, Santa Cruz, CA 95064, USA
}
\date{\today}
\begin{document}
\maketitle
\begin{abstract}
The NGC 5044 galaxy group is dominated by a luminous elliptical galaxy
which is surrounded by $\sim160$ dwarf satellites. The projected
number density profile of this dwarf population deviates within
$\sim1/3$ of the virial radius from a projected NFW-profile, which is
assumed to approximate the underlying total matter distribution. By
means of a semi-analytic model we demonstrate that the interplay
between gravitation, dynamical friction and tidal mass loss and
destruction can explain the observed number density profile. We use
only two parameters in our models: the total to stellar mass
fraction of the satellite halos and the disruption efficiency. The
disruption efficiency is expressed by a minimum radius. If the tidal
radius of a galaxy (halo) falls below this radius it is
assumed to become unobservable. The preferred parameters are an initial
total to stellar mass fraction of $\sim20$ and a disruption radius of
$4\kpc$. In that model about $20$ per cent of all the satellites are totally
disrupted on their orbits within the group environment. 
Dynamical friction is less important in shaping the inner slope
of the number density profile since the reduction in mass by tidal
forces lowers the impact of the friction term. The main destruction
mechanism is tide. In the preferred model the total B-band luminosity
of all disrupted galaxies is about twice the observed luminosity
of the central elliptical galaxy, indicating that a significant fraction of
stars are scattered into the intragroup medium. 
Dwarf galaxy satellites closer to the centre of the NGC 5044 group 
may exhibit optical evidence of partial tidal disruption. 
If dynamical friction
forces the satellite to merge with the central elliptical, the angular
momentum of the satellite tends to be removed at the apo-centre
passage. Afterwards the satellite drops radially towards the
centre.  
\end{abstract}
%%
%%--------------------
\section{Introduction}
%%--------------------
%%
It is generally accepted that the galaxy distribution in rich clusters of
galaxies closely follows the dark matter distribution. In other words the
number density of galaxies observed in clusters is well described by a
Navarro, Frenk and White profile (\citealt{navarro_etal97}). This
behaviour is established by observations (see e.g.~\citealt{
carlberg_etal97, biviano_girardi03, lin_etal04}) as well as 
theoretical investigations (e.g.~\citealt{gao_etal04, kravtsov_etal04,
diaferio_etal01, springel_etal01}). In contradiction to these outcomes
the number density profile of the galaxies in the NGC 5044 group shows
a flattening towards the centre, which is not in agreement with the
standard NFW profile (\citealt{mathews_etal04}). The NGC 5044 group of
galaxies resembles closely the properties of fossil groups
(\citealt{ponman_etal94, jones_etal00, jones_etal03}). The central
elliptical galaxy in NGC 5044, with luminosity $L\B = 4.5 \times
10^{10}L_{B\odot}$ is surrounded by $\sim160$ known dwarf satellites
(\citealt{ferguson_etal90, cellone_buzzoni05}). At a distance of
$33\Mpc$ (e.g. \citealt{tonry_etal01}) the absolute magnitudes of the 
non-central galaxies range from $M\B = -19$ to $-13$,
i.e. $2\times10^7 \lesssim L\B \lesssim 9\times10^9
L_{B\odot}$. About $80$ per cent of the group members appear to be
spheroidal or dE galaxies.

The basic differences between the dark matter substructure 
(subhalo) distribution and the smoothly
distributed component of cosmic matter assemblies is that  
substructures lose mass due to tidal forces and are affected by
dynamical friction, whereas these two mechanisms have no impact on the
smooth component. Both of these mechanisms can cause 
differences between the substructure distribution and the
smoothly distributed matter component in groups and clusters of
galaxies. Thus it is noteworthy that the galaxy number density
profiles in rich clusters are in good agreement with the smooth
density distribution. Dynamical friction and tidal mass loss seem to
balance each other. Both mechanisms acting at the same time result in
a distribution of the substructures which resembles the underlying
smooth distribution. However, the radial number density profile  
of satellite galaxies in the NGC 5044 group does not follow the
underlying matter distribution. We estimate the expected radial
distribution of galaxies in NGC 5044 by following the orbits in the
time dependent potential well of the hosting halo. Since the masses of
the dark halos that initially surround the stellar component of the
satellites are not known well, we treat their {\it total to stellar mass
ratio} ($f\sta = M\tot/M\sta$) as a parameter in our computations. A
second parameter which goes into the integration of the orbits is the
{\it disruption radius} $R\dis$. Satellite halos that are truncated by tides
down to this radius are assumed to reduce below the luminosity
limit of the observations. We will show that the number density
profile of the NGC 5044 galaxy population can be explained by the
action of dynamical friction and tidal forces if total disruption of
galaxies is allowed. When tidal disruption is considered it is no
longer necessary to assume, as \cite{mathews_etal04} have done, that
the radial number density profile of satellite galaxies in NGC 5044 
group requires an inherent deficiency of dwarf satellite galaxies 
at high redshifts. The tidally stripped stars may affect the
intragroup medium, in particular the abundance of heavy elements  
(e.g.~\citealt{zaritsky_etal04}).

This paper is organised as follows. \S~\ref{sec:model} focuses on the
integration of the galaxy orbits within the time dependent potential
well of the hosting halo. In \S~\ref{sec:profiles} the number density
profiles of the resulting satellite distribution at $z=0$ for
different sets of parameters are discussed. \S~\ref{sec:balance}
investigates in detail the impact of dynamical friction versus tidal
disruption. \S~\ref{sec:con} summarises the results obtained.  
%%
%%------------------------------------
\section{Dynamical model for NGC 5044}
\label{sec:model}
%%------------------------------------
The model discussed here is a derivative of the model presented by
\cite{mathews_etal04}. Very recently \cite{zentner_etal05}
demonstrated that this kind of semi-analytical model shows good
agreement with state of 
the art N-body simulations. Similar approaches have been used for a
variety of scientific goals, see e.g.~\cite{bullock_etal00,
bullock_etal01,  zentner_bullock03, koushiappas_etal04,
taylor_babul01, taylor_babul04, islam_etal03, vandenbosch_etal05}. The
local space density of satellite galaxies is assumed to be
proportional to that of the total 
gravitating matter, as if a fixed fraction of baryons formed into
low-luminosity galaxies at a very early time. Thus the number of
galaxies that have entered the virial radius of the host halo
at a given time is proportional to the total amount of mass
accreted at that time. Host and satellite halos are modelled according
to findings of \cite{wechsler_etal2002}. A random B-band luminosity
is assigned to every satellite galaxy in such a way, that their 
luminosity distribution matches the 
distribution observed in NGC 5044. After a galaxy has entered the virial
radius of the NGC 5044 group 
for the first time, its orbit is followed in the 
growing potential well of the host. The integration scheme takes  
dynamical friction and tidal mass loss into account. Satellites are
assumed to be lost if their tidal radius falls below a certain radius.
For simplicity we assume that all stars are formed at redshift 
$z\sta = 7$ (we also employed $z\sta = 3$ and $11$, which does not
change the overall picture). Dark matter halos formed at that epoch have
central densities comparable to that of the stellar densities observed.

The underlying cosmology is the presently favoured concordance
model with $\Om = 0.3$, $\Ol = 0.7$, a Hubble constant of
$70\rm km s^{-1} Mpc^{-1}$ and a normalisation of the power spectrum
of $\sigma_8 = 0.9$.    
%%
%%---------------------------------
\subsection{Modelling the host halo}
%%---------------------------------
%%
Following the analysis of \cite{wechsler_etal2002} the mass growth of
the hosting halo can be described by 
\begin{equation}
\label{equ:evoma}
M_v(z) = M_{\rm v,0}e^{-2a_fz}
\end{equation}
where $a_f = 1/(1+z_f)$ defines the halo formation time. 
The expansion factor $a_f$ at the time of formation for NGC 5044 can
be estimated by
\begin{equation}
\label{equ:evoco}	
c\vir = {c_1 a_0\over a_f}\ .
\end{equation}
If at present time $a_0=1$ a concentration of $c\vir = 11.1$ is
assumed and $c_1$ is set to $5.125$ (\citealt{zentner_etal05}) the
expansion factor at formation turns out to be $0.46$ which translates
into a formation redshift of $z_f = 1.2$. Knowing the evolution of
mass (Eq.~\ref{equ:evoma}) and concentration (Eq.~\ref{equ:evoco}) the
virial radius at that redshift can be determined by the following
formula.  
\begin{equation}
\label{equ:evorv}
r\vir = \left({3M(a)\over4\pi}{1\over\Delta(a)\rho_c(a)}\right)^{1/3} 
\end{equation}
where $\Delta(a)$ is a outcome of the spherical collapse approximation
(see \citealt{eke_etal98, bryan_norman98}). It gives the virialised
overdensity with respect to the critical density $\rho_c(a)$ at that
epoch. $M(a)$ is the time dependent mass specified in
Eq.~\ref{equ:evoma}. With the above recipe the dark matter profile 
and the gravitational potential of the host can be computed for any
radius and redshift. The orbits of the satellite halos are computed
within this potential well.
%%
%%------------------------------------
\subsection{Initial galaxy properties}
%%------------------------------------
%%
We assume that all satellite galaxies form at the same early redshift
$z\sta$ and that their space density is proportional to 
that of the dark matter. 
Thus the satellites enter the host halo at the same rate as the dark
matter. With this simple model the host mass $M_{\rm v,i}$ when the ith
satellite first entered the virial radius of the 
host can be found from 
\begin{equation}
M_{\rm v,i} = \mathcal{R}_M M_{\rm v,0}
\end{equation}
where $0\leq\mathcal{R}_M\leq1$ is a random number 
and $M_{\rm v,0} = 3.9 \times 10^{13}$ $M_{\odot}$ is the 
current virial mass of the NGC 5044 group. The virial
mass of the host at entry of the ith satellite can be translated into
a redshift $z_i$ via Eq.~\ref{equ:evoma} and a virial radius $r_{\rm v,i}$
by Eq.~\ref{equ:evorv}. The velocity of the galaxy at the virial
radius is the free-fall velocity from the turnaround radius
$r_t=2r_{\rm v,i}$. 
\begin{equation}
\label{equ:vinit}
u_{\rm v,i}= \left(\G M_{\rm v,i}/r_{\rm v,i}\right)^{1/2}
\end{equation}
The stellar mass of the ith galaxy is 
\begin{equation}
\label{equ:star}
m_{\ast,i} = \Upsilon\B L_{B,i}\ . 
\end{equation}
Here the galactic luminosity is randomly selected from the power law
part of the Schechter luminosity function $dN/dL\B \propto L\B^{-1.2}$
(which is satisfied in NGC 5044): 
\begin{equation}
L_{B,i} =  
\left[L_1^{-0.2}-\mathcal{R}_L(L_1^{-0.2}-L_2^{-0.2})\right]^{-5}\ , 
\end{equation}
with $L_1 = 2\times10^7 L_{B\odot}$ and $L_2 = 9\times10^9
L_{B\odot}$ and $\mathcal{R}_L$ is another random
($0\leq\mathcal{R}_M\leq1$). For the stellar mass to light ratio 
$\Upsilon\B$ in Eq.~\ref{equ:star} we adopt 
\begin{equation}
\Upsilon_{\!B} = 0.71\times L^{0.1}
\end{equation}
as discussed by \cite{trujillo_etal04}. We assume that the galaxies
form in isolation. Thus it is reasonable to embed them into a dark
matter halo. Since the ratio between total and stellar 
matter in isolated dwarf galaxies is not well determined, we
introduce a parameter $f\sta=M\tot/M\sta$ which allows us to vary the
star formation efficiency. The total mass of the satellite including
the dark matter halo is then given by  
\begin{equation}
m_{tot,i} = f\sta \Upsilon\B L_{\!B,i}.
\end{equation}
The density profile of all matter components  - dark, stellar and
gaseous -  is assumed to follow a NFW-profile which is in agreement 
with recent studies by \cite{colin_etal04} who showed that the
density profiles of dwarf (N-body) halos are well fitted by
NFW-profiles. The concentration of each satellite halo accreting at
redshift $z_i$ is computed by means of Eq.~\ref{equ:evoco}. For this
purpose $a_0$ has to be interpreted as the expansion factor at arrival
($a_0 = 1/(1+z_i)$) and $a_f$ corresponds to the formation redshift of
the halo $a_f=1/(1+z\sta)$. As mentioned above, the choice of  
different formation redshifts $z\sta= 3, 7, 11$, 
has marginal influence on the results. Subsequently, we assume
$z\sta=7$ for the global formation redshift of the satellite halos.
This simplifications do not pay tribute to the
dissipative baryonic processes, like radiation or adiabatic
contraction (see e.g.~\citealt{gnedin_etal04}). Due 
to these processes the central parts of the satellite halos might be
more tightly bound and the total tidal destruction
might become more difficult. This shortcoming can be compensated by the
variation of the the second parameter introduced here, the {\it
disruption radius} $R\dis$. If the tidal forces manage to strip the outer
matter shells down to this radius, the satellite galaxy is assumed to
vanish, i.e. fall below optical observability. We adjust this parameter to
reach agreement with the global satellite distribution observed in NGC
5044, without tackling the complex processes close to the halo centre.
%%
%%------------------------------------
\subsection{Integration of the orbits}
%%------------------------------------
%%
We integrate the orbits of each satellite galaxy starting at 
the virial radius when the satellite first enters the 
virialised halo of the host. The absolute value of the initial
velocity is  computed under the assumption that the satellite has
experienced free-fall from twice the virial radius, 
Eq.~\ref{equ:vinit}. The absolute value of the velocity is
distributed to a radial and a tangential 
component. The tangential component is assigned in
accordance with the results by \cite{tormen_97} (see also
\citealt{zentner_etal05}). He gives a 
distribution of circularity $\epsilon = J/J_c$ of dark matter
particles at the virial radius. Here $J$ is the specific angular
momentum of the dark matter particle and $J_c$ is the maximum 
possible specific angular momentum of bound particles at that radius.
$\epsilon = 0$ or 1 indicates radial or circular orbits, respectively.  
Based on these results we model (see \citealt{mathews_etal04}) the
probability density for circularity $p^{\prime}=dp/d\epsilon$ with a
third order polynomial, subject to the constraints that
$p^{\prime}(0)=p^{\prime}(1)=0$ and normalised so the integral of
$p^{\prime}(\epsilon)$ from 0 to 1 is unity. The resulting
distribution is 
\begin{equation}
p^{\prime}(\epsilon)=a\epsilon^3-(1.5a+6)\epsilon^2+(0.5a+6)\epsilon
\end{equation} 
where 
\begin{equation}
a = 60 ( 1 - 2 \langle\epsilon\rangle).
\end{equation} 
The results of Tormen suggest that $\langle\epsilon\rangle\approx0.5$,
for which $p^{\prime}(\epsilon)$ becomes a symmetric quadratic.

The orbits of the satellites are found by solving 
\begin{equation}
{d\vec{r}\over dt} = \vec{u} 
\hspace{0.1\hsize}\text{and}\hspace{0.1\hsize}
{d\vec{u}\over dt} = - {\G M(r)\over r^2}{\vec{r}\over r} +
\left({d\vec{u}\over dt}\right)_{df} \ , 
\end{equation} 
where $M(r)$ is the NFW mass within the radius $r$ of the host halo at
that time. Orbits are computed from redshift $z_i$ (redshift of entry)
to $z = 0$. The deceleration by dynamical friction
(\citealt{chandrasekhar_43,colpi_etal99}) can be described by 
\begin{equation}
\left({d\vec{u}\over dt}\right)_{df} = 
-\vec{u} 4 \pi \ln\Lambda \G^2 m \rho
u^{-3}[{\rm erf}(X)-{2\over\pi^{1/2}}Xe^{-X^2}]\ . 
\end{equation}
Here $\rho(r,t)$ is the local density of the host halo, 
$u=|\vec{u}|$ is the velocity of the satellite, $X =
u/(\sqrt{2}\sigma)$ and $\sigma(r,t)$ is the mean velocity dispersion
in NGC 5044. We assume that the total dispersion can be approximated
by the cold dark matter dispersion presented by \cite{hoeft_etal04}
(see also \citealt{mathews_etal04}). We choose $\ln\Lambda = 3$ as
suggested by \cite{zhang_etal02} which is the point mass  
approximation. This approximation is justified by the very efficient
tidal mass stripping which effectively truncates the radius of the
satellite halo.  

The tidal truncation radius is estimated by the Jacobi limit
(see \citealt{binney_tremaine87,hayashi_etal2004}). 
\begin{equation}
r_J = \left({m_{tot,i}\over3M(<D)}\right)^{1/3}D
\end{equation}
Here $D$ is the distance of the satellite to the centre of the group,
$M(<D)$ is the mass of the host within $D$ and $m_{tot,i}$ is the
current mass of the halo. If the resulting truncation radius $r_J$ is
smaller than the current radius of the satellite halo, 
then the truncation radius is assumed
to be the current halo radius and all the matter outside of $r_J$ is
stripped instantaneously. In our model a minimum disruption radius
$R\dis$ is introduced. If the truncation radius falls below this
radius the satellite's luminosity is assumed to fall below
observability. We will show models for $R\dis = 0,2,4$ and $6\kpc$,
respectively. For $R\dis = 0$ all the satellites survive. For
comparison the effective radius of the lowest luminosity 
dwarfs in NGC 5044 is approximately $4\kpc$. We do not allow a
change of the density or velocity profile of the satellite halo 
following the truncation. \cite{kazantzidis_etal04} have shown
that a cuspy density profile remains after mass is lost by tidal
forces. However, tidal heating (see e.g. \citealt{gnedin_03}) may
reduce the central density and therefore reduce the mass contained
within the truncation radius. In fact by means of N-body simulations
\cite{hayashi_etal2004} and \cite{kravtsov_etal04} have shown that some
mass is lost even within the tidal radius of sub-halos orbiting in the
potential well of a host halo. \cite{hayashi_etal2004} compare
different estimators for the truncation radius, concluding that the
Jacoby limit most accurately estimates this additional tidal mass
loss. Therefore we apply the Jacoby limit. Nevertheless,  
due to the dense baryonic cores, more mass may be retained than
estimated from pure N-body investigations. We assume that the net trend of
the different processes can be represented by an appropriate choice of the
disruption radius $R\dis$ that generates the final observed number
distribution. Since the force exerted by dynamical friction is
proportional to the mass squared, tidal mass loss strongly reduces the
impact of dynamical friction.
\begin{figure*}
\epsfig{file=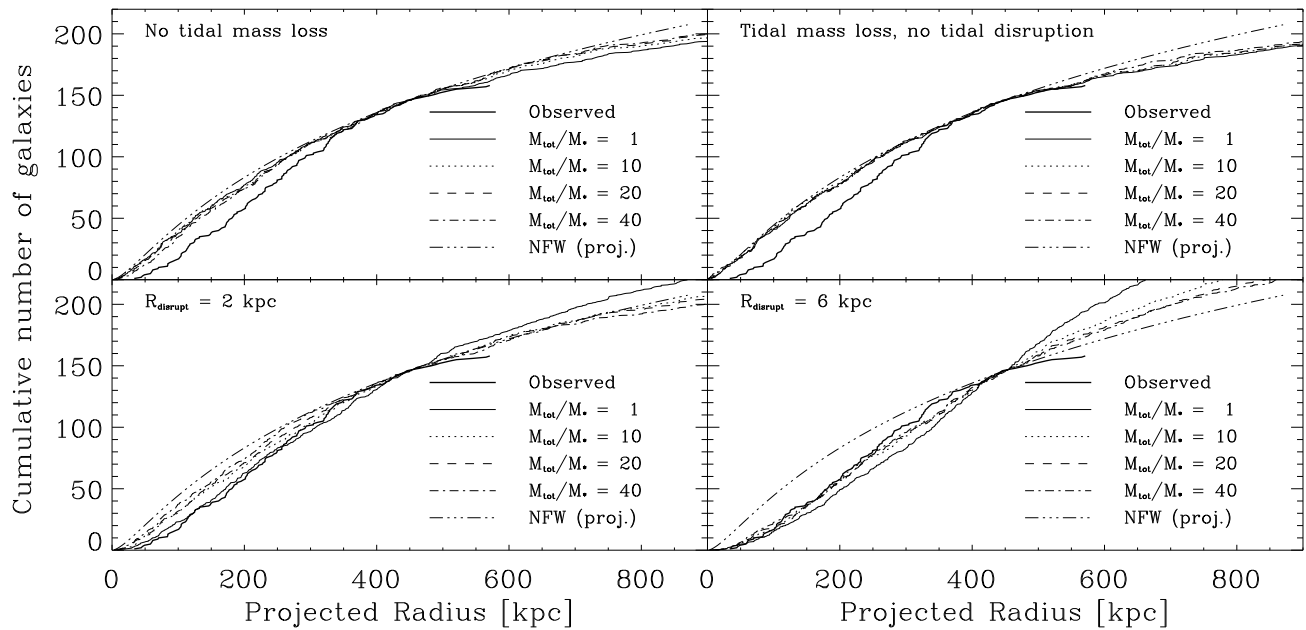,width=\hsize}  
\caption{\label{fig:zfor07}
Comparison of the observed galaxy distribution in NGC 5044 (thick solid
line) with the different model profiles. The various thin lines
indicate alternate mass ratios between the total mass of the halo and
the stellar component as indicated by the legend within each plot. The
upper left-hand panel shows the resulting profiles if tidal mass loss is
not taken into account. In the upper right panel tidal mass loss is
tracked but the halos never are assumed to be disrupted. The lower
panels depict the profiles if halo disruption is implemented. The
disruption radii are assumed to be on left- and right-hand side $2\kpc$
and $6\kpc$, respectively. 
}
\end{figure*}
%%
%%-------------------------------
\section{Number density profiles}
\label{sec:profiles}
%%-------------------------------
%%
%%
\begin{figure*}
\epsfig{file=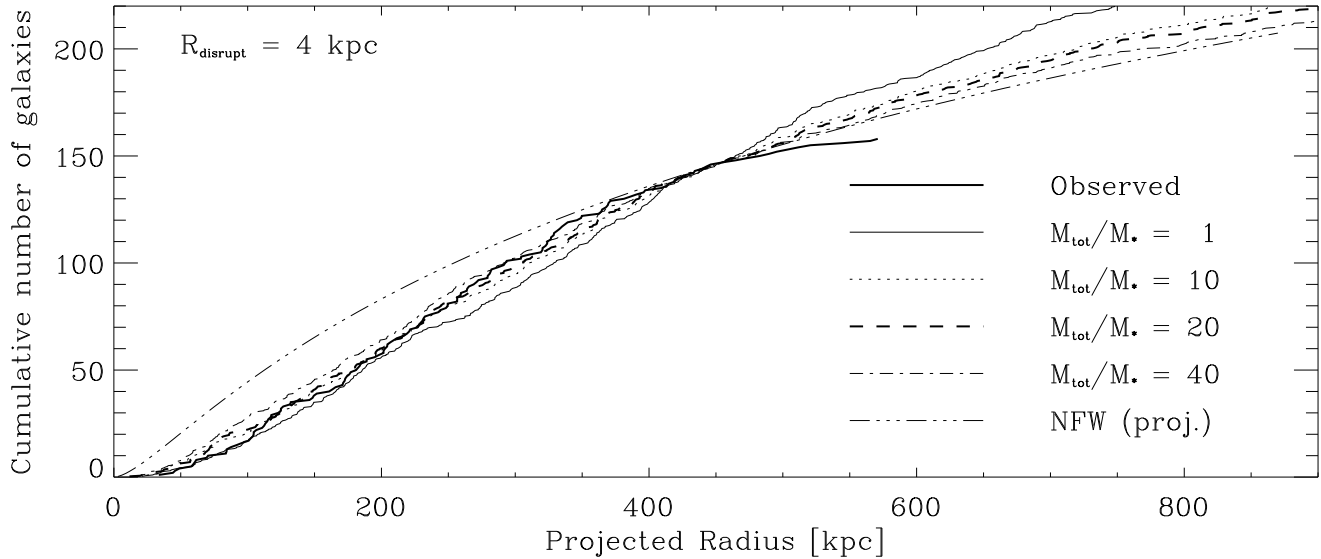,width=\hsize}
\caption{\label{fig:zfor07Rup004}
Comparison of the observed galaxy distribution in NGC 5044 (thick solid
line) with the different model profiles for a disruption radius of
$4\kpc$. The various thin lines indicate alternate mass ratios between
the total mass of the halo and the stellar component as indicated by
the legend. The thick dashed line belonging to a stellar to total mass
ratio of 20 marks the favoured model.   
}
\end{figure*}

In this section we present the resulting number density profiles for
different sets of parameters. As mentioned above we use only two
parameters, firstly the initial total to stellar mass ratio $f\sta$
to vary the contribution of the cold dark matter halo, secondly the
disruption radius $R\dis$ which modulates the destruction rate due to
tidal forces. Fig.~\ref{fig:zfor07} shows the results for four
different tidal disruption/destruction efficiencies. 

The upper left panel 
displays the radial distribution resulting from dynamical
friction without tidal mass loss. This scenario is rather unphysical
since observations (e.g.\citealt{majewski_etal04, rocha-pinto_etal04})
and N-body simulations (e.g.~\citealt{diemand_etal04, kravtsov_etal04,
nagai_kravtsov04}) have proven that tidal mass loss is
significant. However it is interesting to note that 
the different masses of the satellite halos due to different
$f\sta$-values do not result in greatly different profiles. All
profiles fall a little below the projected NFW-profile and are
equally discrepant when compared to the observations. Dynamical friction
shifts the whole distribution inwards without seriously affecting 
the overall shape.

The upper panel on the right in Fig.~\ref{fig:zfor07} shows
the resulting distribution if tidal mass loss is allowed but total
tidal destruction is not allowed ($R_{dis} = 0$). 
This prescription is similar to the
treatment of the cluster galaxies in \cite{gao_etal04}. In agreement
with their findings the galaxy number density profile follows closely
the NFW-profile (at least within the $\sim{2\over3}$ of the virial
radius). However the resulting profiles are not in agreement with
observations of NGC 5044. 
This difference emphasises 
that the shallow potential wells of the faint dwarf galaxies 
in NGC 5044 must be considered.
Observations addressing the dwarf populations in
clusters of galaxies suggest that the dwarf galaxies can be
strongly affected by tidal forces. According to \cite{dahlen_etal04}
and \cite{lin_etal04} the decreasing dwarf-to-giant ratio with
increasing surface density in clusters of galaxies indicates that the
high density environments are hostile to dwarf galaxies.  

The lower panels in Fig.~\ref{fig:zfor07} compare the 
the observed projected satellite distribution with model
profiles for two different tidal destruction efficiencies, 
$R\dis = 2$ and $R\dis = 6$ kpc. In the lower left-hand
panel only the distribution with $f\sta = 1$ matches the observed
distribution below $450\kpc$. But the strong increase above $450\kpc$
and the fact that a mass to light ratio of 1 entirely neglects the
contribution of cold dark matter gives evidence against this
particular model. The model profiles in the lower right-hand panel agree
with the observation for $f\sta = 10,20$ and $40$ below radii of
$450\kpc$, but obviously exceed the data and the NFW profile for large
radii. (It should be mentioned that the observed distribution is
likely to be incomplete at large radii.) 
The census of surviving satellite galaxies 
in Tab.~\ref{tab:celu} indicates that
a large fraction of satellites ($\sim30\text{ per cent}$) have been destroyed. 

Fig.~\ref{fig:zfor07Rup004} presents a model with 
disruption radius $R\dis=4\kpc$ that matches the data for NGC 5044
very well. Interestingly this $R\dis$ turns out to be equal to the
effective radius of the faintest galaxies observed in NGC 5044, if the
minimum B-band luminosity $L\B=2\times10^7L_{\!B\odot}$ is
transformed into an effective radius $R_e$ using the relation given in
\cite{derijcke_etal04}. The total to stellar mass 
ratio which approximates the
data best is $f\sta=20$, or perhaps a bit lower since choosing $f\sta=10$
results in very a similar behaviour. \cite{prada_etal03} have shown
that the consideration of the dark matter halo of isolated galaxies
leads to $M\tot/L\B\approx130$ for galaxies within the luminosity
range of  $1\times10^{10}\lesssim L\B\lesssim 4\times10^{10}
L_{\!B\odot}$ corresponding roughly to $f_* = M\tot/M\sta = 16$.
Therefore the chosen parameter set seems to be in good agreement with
recent observations. As indicated by the bold line in
Tab.~\ref{tab:celu}, these parameters lead to a 
stellar loss rate of $\sim20$ per cent. The
total luminosity of all these lost galaxies is 
$9.4\times10^{10}L_{\!B\odot}$. This luminosity is twice as high as
the luminosity of the central elliptical galaxy in NGC 5044, 
$L_B = 4.5 \times 10^{10}$ $L_{B\odot}$, 
but only $\sim35$ per cent of the total B-band luminosity 
in the group including all dwarf galaxies 
$L_{B,\rm d} \approx 13.7 \times 10^{10}$ $L_{B\odot}$ 
(corrected for incompleteness, the corresponding B-band luminosity
within $r_{vir}$ in the best fitting model is 
$L_{B,\rm d} \approx 16.9 \times 10^{10}$ 
$L_{B\odot}$) and galaxies disrupted in the model. Evidently, the
missing stars are scattered in the intragroup medium (see
e.g. \citealt{zaritsky_etal04, sommer-larsen_etal05}).  

\begin{table}
\begin{center}
\begin{tabular}{cccccc}
\hline
$R\dis$& $M\tot/M\sta$& $N_{cen}/N_{vir}$&$N_{fric}$& $N_{tid}$&
$L_{B\odot}$\\\hline\hline  
2 &  1 & 0.26 & 0 &  173 & $6.4\times10^{10}$\\
2 & 10 & 0.10 & 0 &   63 & $3.8\times10^{10}$\\
2 & 20 & 0.08 & 2 &   47 & $4.9\times10^{10}$\\
2 & 40 & 0.06 & 8 &   31 & $6.5\times10^{10}$\\\hline
4 &  1 & 0.68 & 0 &  486 & $2.4\times10^{11}$\\
4 & 10 & 0.25 & 0 &  170 & $8.4\times10^{10}$\\
{\bf4} & {\bf20} & {\bf0.21} & {\bf3} &  {\bf137} & ${\bf9.4\times10^{10}}$\\
4 & 40 & 0.18 & 7 &  110 & $1.1\times10^{11}$\\\hline
6 &  1 & 1.34 & 0 & 1075 & $6.3\times10^{11}$\\
6 & 10 & 0.42 & 0 &  297 & $1.5\times10^{11}$\\
6 & 20 & 0.32 & 3 &  219 & $1.3\times10^{11}$\\
6 & 40 & 0.28 & 7 &  179 & $1.4\times10^{11}$
\end{tabular}
\caption{\label{tab:celu}
Survival rates. The first two columns display the 
disruption radius and total to stellar mass, respectively. The third
column gives the fraction of galaxies that ended in the centre to the
current number of galaxies within the virial radius. The following two
columns list the number of galaxies that disappeared due to
dynamical friction and tidal disruption. The last column shows the
summed luminosity of all galaxies that have been destroyed. The bold
face line indicates the favoured model.}
\end{center}
\end{table}

We have shown the results for a uniform formation redshift of $z\sta =
7$. Different formation redshifts result in different concentrations
of the satellite halos and thus might affect the tidal stripping
efficiency. Therefore we have redone the modelling with formation
redshifts of $z\sta = 3$ and $11$. The differences are marginal. We
conclude this section by repeating that the model parameters
$f\sta=20$, $R\dis=4\kpc$ and $z\sta = 7$ match the data very well.  
%%
%%----------------------------------------------
\section{Dynamical friction and tidal mass loss}
\label{sec:balance}
%%----------------------------------------------
%%
\begin{figure}
\epsfig{file=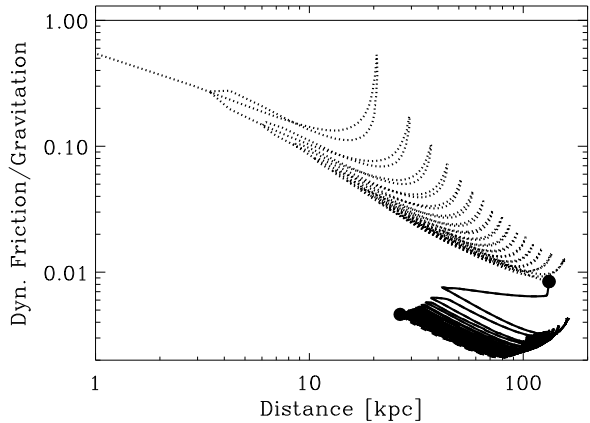,width=0.95\hsize}  
\caption{\label{fig:id00071}
Trajectory of a model galaxy. The ratio of forces caused by
dynamical friction and by gravitation is ploted versus the 
physical distance of
the galaxy from the centre of the group. The large, right dot marks the
first penetration of the galaxy through the virial radius. 
The dotted line represents the path of the galaxy if tidal mass loss
is suppressed but retaining dynamical friction. 
The solid and dashed (almost totally hidden) lines show the path if
tidal mass loss is also taken into account. The left dot marks the
location where the galaxy is assumed to be tidally lost if a disruption
radius of $4\kpc$ is applied. The dashed line follows the path of the
galaxy till $z=0$ if the galaxy suffers no tidal
destruction ($R\dis = 0$). 
The solid horizontal line at the top separates the
friction dominated area from the area where gravitation prevails.} 
\end{figure}
\begin{figure}
\epsfig{file=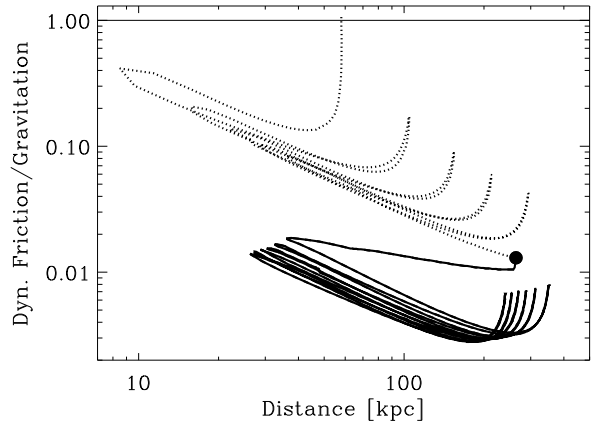,width=0.95\hsize}  
\caption{\label{fig:id00025}
Trajectory of a model galaxy, the description is the same as
in Fig.~\ref{fig:id00071}. The absence of the second black dot 
indicates that this halo is not destroyed by tidal
forces when a disruption radius of $4\kpc$ is applied.} 
\end{figure}
\begin{figure}
\epsfig{file=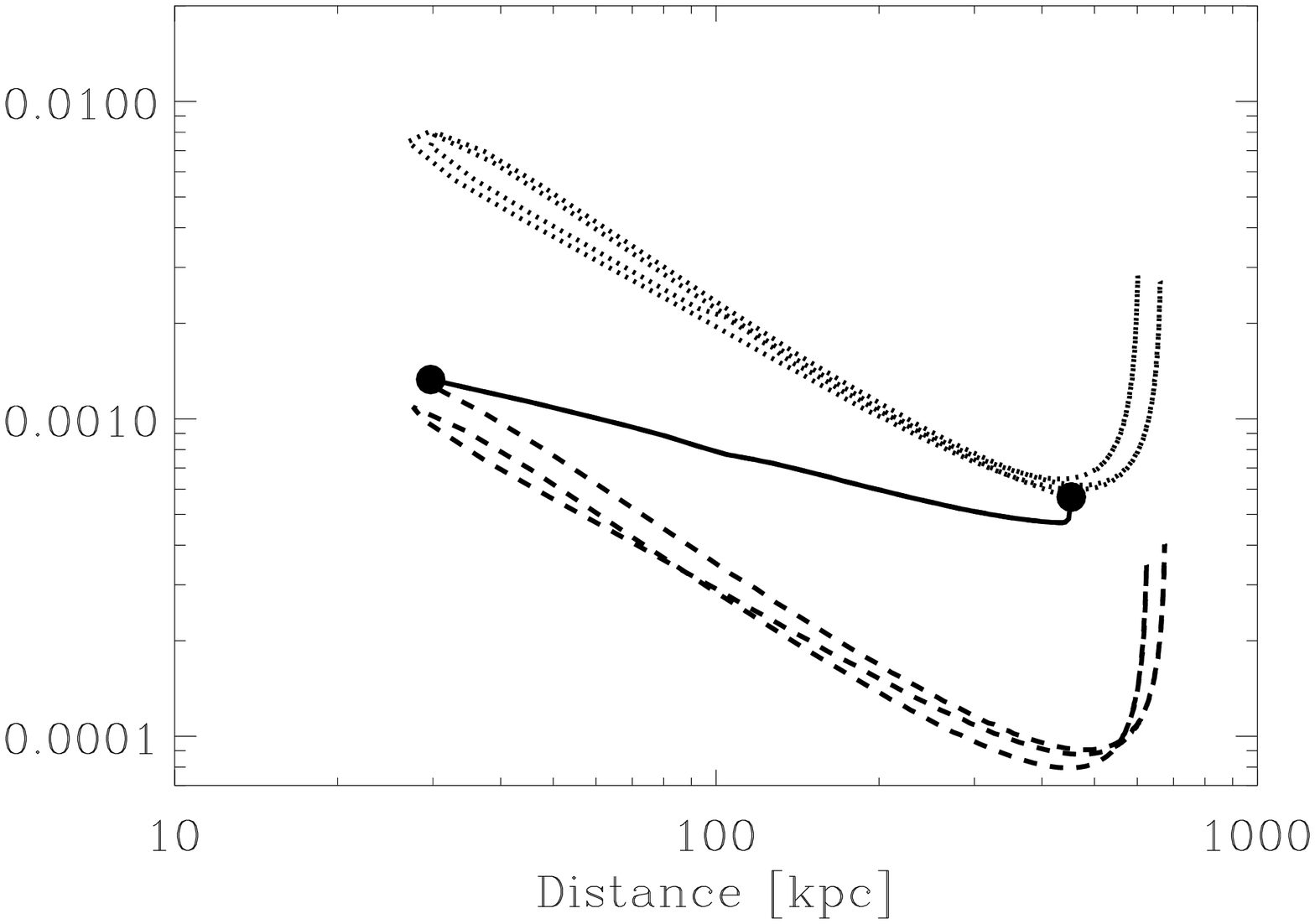,width=0.95\hsize}  
\caption{\label{fig:id00044}
Trajectory of a model galaxy, the description is the same as
in Fig.~\ref{fig:id00071}.}
\end{figure}

In this section we show the orbital behaviour of three particular
galaxies. These three cases are chosen to make some of the features
discussed above more explicit. In Figs.~\ref{fig:id00071},
\ref{fig:id00025} and ~\ref{fig:id00044} we compare the
behaviour of satellite halos having the same initial conditions with
(solid and dashed lines) and without (dotted lines) tidal mass
loss. The dashed lines indicate the orbits without tidal destruction,
whereas the solid line ends (additionally marked by a second large
dot) when the truncation radius falls below $4\kpc$. 
In these figures the ratio of the forces due to dynamical friction and
gravitation ($F/G$) is plotted versus the distance of the satellite to the
centre of the group. The horizontal line for $F/G = 1$ indicates the
threshold where dynamical friction equals the force exerted by
gravitation. Once an orbit transgresses above this line, it is dominated
by dynamical friction and the galaxy drops nearly radially toward 
the centre.

The large, right dot in Fig.~\ref{fig:id00071} marks the 
first passage through the 
virial radius and therefore the beginning of integration. After
orbiting once around the group centre 
the satellite moves farther outwards bacause its
initial velocity was the free-fall velocity from
twice the virial radius. The eccentricity of the orbit
without tidal stripping (dotted line) changes from initially $\epsilon
\approx 0.6$ to a final value of $\approx 0.7$. Interestingly the
contribution of friction to the total force shows a minimum for
intermediate radii and rises again if the satellite moves further
out. Dynamical friction is not negligible during the apo-centre
passage. For small velocities $F/G$ is roughly proportional to $r/u$,
where $r$ is the distance to the group centre and $u$ is absolute 
values of the velocity. Frictional decay of the 
tangential velocity at apo-centre leads to more radial orbits, if
tidal mass loss is not taken into account. Furthermore the computed
orbit indicates that satellites can first transgress the $F/G=1$ line
during their apo-centre passage. In that case all the angular momentum
of the satellite  
is lost and it approaches the centre on a pure radial orbit. 
Such an orbit might be appropriate for a massive black hole 
that experiences dynamical friction without tidal losses. 
The eccentricity of the orbit with tidal stripping (solid/dashed line)
stays $\approx 0.6$ all the time. This is due to that fact that tidal
stripping reduces the action of dynamical friction. Most of the mass
is lost during the first infall, as can be inferred by the shallow
slope till the first peri-centre passage. The large, left dot indicates
where the tidal destruction is assumed by our approach. Here $R\dis =
4\kpc$ was chosen. Fig.~\ref{fig:id00071} shows that after $\sim 10$ 
orbits the satellite is destroyed at a distance $\sim 25\kpc$ 
from the centre, which is quite close to the central elliptical galaxy.

The orbit shown in Fig.~\ref{fig:id00025} has a similar behaviour, 
but differs from the previous orbit in that 
the satellite is not destroyed 
by tidal forces (no second large dot). 
Without tidal stripping and mass loss (dotted line) 
the satellite halo eventually falls toward the group centre 
as indicated by the final passage above the $F/G=1$ line. 
This catastrophic loss of angular momentum by 
dynamical friction occurs during an apo-centre passage. 
When all its orbital angular momentum is lost, the satellite 
approaches the centre on a radial orbit starting at $\sim60\kpc$ from
the centre. The eccentricity of the orbit without tidal stripping
changes from $\approx 0.8$ to $\approx 0.7$, whereas the orbit with
tidal mass loss retains its initially eccentricity of $\approx 0.8$. 

Fig.~\ref{fig:id00044} is an example for a very radial orbit in which 
the eccentricity remains at 
$\approx0.9$ for all orbits, either with or without 
tidal mass loss. The large initial radius $\sim450\kpc$
indicates that the satellite approaches rather late, that means the
potential well of the hosting group is already very deep. The large,
left dot marks the destruction during the first peri-centre passage,
if tidal stripping and destruction are allowed ($R\dis=4\kpc$). For
this particular case, which is rather extreme, the fast destruction
seems reliable, since the satellite is falling from  $\sim900\kpc$
(twice the virial radius at epoch of entry) down to $\sim20\kpc$ which
is very close to the central elliptical galaxy.

The detailed analysis of the individual orbits 
clearly demonstrates that the
influence of dynamical friction is largely reduced by tidal mass
loss. The estimation of the tidal radius by the Jacobi limit and the
instantaneous stripping up to this radius lead to a rapid decrease of
the halo mass during the first infall. The instantaneous stripping
description might reduce the mass a bit to fast, but after very few
orbits the Jacobi radius should be a good approximation. On subsequent
orbits dynamical friction tends to push the halo closer to the centre
which is accompanied by further tidal mass loss, since the tidal
radius of the halo is proportional to its distance from the
centre. The force exerted by dynamical friction is proportional to the
halo mass squared, thus the loss of mass reduces the impact of
friction further. In conclusion the dominant process for shaping the  
inner number density profile of groups like NGC 5044 
is tidal disruption and/or destruction.   
%%
%%------------------
\section{Conclusion}
\label{sec:con}
%%------------------
%%
Using a simple model we find that the interplay between dynamical
friction and tidal disruption can explain the shallow slope of the 
projected number density profiles of
satellite galaxies of the fossil group NGC 5044. 
The strong deviation of observed satellite galaxies 
from NFW profiles is a result of the 
tidal disruption of the stellar cores of galaxies orbiting closer 
to the group centre. 
This feature may be more naturally explained in this manner than 
by assuming an absence of dwarf galaxies at high redshifts
(\citealt{mathews_etal04}).
We use two parameters to
model the observed dwarf galaxy distribution: the total to
stellar mass ratio $f\sta$ and the disruption radius $R\dis$. With 
parameters $f\sta\approx20$ and $R\dis\approx 4\kpc$ the model
can match the observed projected radial number distribution 
in NGC 5044 
very well.  $f\sta\approx20$ is in good agreement with the
findings of \cite{prada_etal03}. They show that the $M\tot/L\B$ of
isolated galaxies with B-band luminosities of $1\times10^{10}\lesssim
L\B\lesssim 4\times10^{10} L_{\!B\odot}$ is in the range between 100
and 150. For $f\sta=20$  we obtain $M\tot/L\B\approx160$. 
If we use the $R_e-L\B$ relation presented by \cite{derijcke_etal04}
to compute the effective radius for our lowest luminosity model
galaxies ($L\B\approx2\times10^{7}L_{\!B\odot}$) we obtain
$R_e\approx4\kpc$. Therefore, the global galaxy distribution in NGC
5044 seems to favour a destruction radius that is close to the
minimal $R_e$ of the group satellites. 

We find that the total luminosity of the tidally disrupted 
galaxies is about twice the luminosity of the central elliptical 
or $\sim 35 \text{ per cent}$ of the total optical light of the group, allowing 
for incompleteness of identified satellite galaxies. This indicates
that a large fraction of satellite stars are scattered into the
intragroup medium by tidal forces. Using the near-infrared $K$-band 
properties of clusters \cite{lin_mohr04} estimate that a large 
fraction ($50 \text{ per cent}$)  of the total stellar luminosity is contributed by
intracluster light. This fraction increases with cluster mass. 
\cite{naill_etal05} searched the Fornax cluster for novae
in between the galaxies and find that $\sim16-41 \text{ per cent}$ of the total
cluster light comes from the intracluster stars.  
Dwarf galaxies orbiting close to the centre of groups like 
NGC 5044 may show morphological evidence of partial tidal 
destruction.
\section*{Acknowledgements}
The authors would like to thank the anonymous referee for his
comments which helped to clarify the text. This work has been
supported by NSF grant AST 00-98351 and NASA grant NAG5-13275 for
which we are very grateful.  
%%----------------
%%\bibliography{/home/fal/draft/dy5044/lit}
%%----------------

%%----------------
\end{document}